\documentclass[aps,pre,twocolumn,showpacs,floatfix,preprintnumbers,amsmath,amssymb]{revtex4-1}
\usepackage{epsfig}
\usepackage{graphicx}
\usepackage{dcolumn}
\usepackage{bm}

\begin{document}
\title{Record statistics of financial time series and geometric random walks}
\author{Behlool Sabir}
\altaffiliation[Present Address : ]{InvenZone, Sakinaka, Mumbai 400072, India.}
\author{M. S. Santhanam}
\affiliation{Indian Institute of Science Education and Research,\\
Dr. Homi Bhabha Road, Pune 411 008, India.}
\date{\today}

\begin{abstract}
The study of record statistics of correlated series is
gaining momentum. In this work, we
study the records statistics of the time series of select stock market data
and the geometric random walk, primarily through simulations. We show that
the distribution of the age of records is a power law with
the exponent $\alpha$ lying in the range $1.5 \le \alpha \le 1.8$.
Further, the longest record ages follow the Fr\'{e}chet distribution
of extreme value theory. The records statistics of geometric random walk series is in good
agreement with that from the empirical stock data.
\end{abstract}
\maketitle

\section{Introduction}
  In popular parlance, records are associated with record breaking
events. Common examples include extreme weather events such as the
occurrence of lowest or highest temperatures \cite{wri},
unparalleled sport performances
in Olympic and other events \cite{ubolt}, financial downturns like
the major stock market crashes. In recent years, there
is an increasing interest in the study of record statistics
in the context of global warming and climate change \cite{rc},
occurrence of cyclones and floods \cite{pson} and stock markets. 
In physics, records statistics is useful in understanding
the behavior of stochastic motion of a domain wall in metallic ferromagnetic
materials \cite{dwd} and as an alternative indicator of quantum chaos in
kicked rotor model \cite{sas}. Even as the record breaking events continue to
enjoy media attention, there is also an increased research interest in
the statistical study of record events \cite{maj1,recs,temp,recs-rw,mrw,recs-rw1}.

For a discretely sampled stochastic time series $x_t, t=1,2,3....N$, record
events are those that are larger (smaller) than all the preceding events.
An event at $t=T$ would be an upper record if $x_T > \mbox{max}(x_1,x_2,...x_{T-1})$.
Then, some of the relevant questions of interest are the probability for the occurrence
of record at any given time, mean number of records in a certain time window
and record age, {\it i.e}, how long a record is expected to survive. The result for
most of these questions is known for uncorrelated random variables \cite{sz}.
However, it is known that most of the physically observed time series, e.g, temperature,
stock market volatility, earthquake magnitudes, are strongly
correlated \cite{bunde}. The record statistics for such cases is beginning to receive
research attention.

Recently, the record statistics of
correlated series such as the positions of random walker was studied \cite{maj1,recs-rw,mrw}.
Random walk is a fundamental model in physics and has applications 
in many areas including the dynamics of stock markets.
It was show that if the increments of the random walker are drawn from a  continuous and symmetric function
$\phi(\xi)$, then
the mean number of records, for large $N$, is proportional to $\sqrt{N}$
and the mean record age $\langle r \rangle \propto N$ \cite{maj1}.
These results have further been generalized to the case of random walk with a constant drift
with application to stock market data \cite{wer} and also to multiple random walkers \cite{mrw}.

Inspite of such growing interest in correlated series, very few works have focussed
on the records statistics in empirical stock data \cite{mrw, recs-rw1,recs-fin}.
In this paper, we report on the record statistics of empirical stock data to understand
two quantities of interest not studied earlier, namely, {\sl (i)} the distribution of record age and
{\sl (ii)} the distribution of longest record ages. We present our analysis of stock data 
in the context of geometric random walk model, which is considered as one of the
suitable models for the dynamics of stock data \cite{grw-lr}. In addition, it must be pointed out
that GRW has other applications as well, including as a model for interacting neurons \cite{kuhn}.

In this paper, we analyze the upper record statistics for 18 stocks, for which longest data is available
in the public domain. The data used in this work is described in the Appendix.
Most aspects of record statistics, especially quantities
such as the mean number of records, record age distribution,
longest record age etc., depend only on the
position of record breaking event on time axis and {\sl not} on its
magnitude. We study these quantities using geometric random walk as
the benchmark model. We show that both for the records in stock data and geometric random walk series the 
distribution of record age $r$ is consistent with $P(r) \sim r^{-\alpha}$, with the exponent
$1.5 \le \alpha \le 1.8$ and the
longest records $r_{max}$ fall in the class of type-II generalized extreme value (Fr\'{e}chet) distribution.

\section{Distribution of record ages}
Geometric random walk (GRW) has not attracted as much
attention as the random walk model except in the context of
financial applications \cite{grw-lr}. GRW model is given by,
\begin{equation}
y_{i+1} = y_i ~ \exp(\xi_i), \;\;\;\;\;\;\; i=1,2,3 \dots N.
\label{grw}
\end{equation}
In this, $\xi_i$ is Gaussian distributed $G(\mu,\sigma)$ with mean $\mu$
and standard deviation $\sigma$. This
implies that the 'log returns' $R_i=\log(y_{i+1}/y_i)$ are also Gaussian.
The log-returns from the empirical stock data is known to be approximately
Gaussian distributed over a wide range of timescales \cite{grw-lr}.

Record age is the time duration $r$ between two successive occurrences of
a record, {\it i.e}, the time for which a record survives. Record age distribution
will provide insights into how long a record can be expected to survive and
is useful in hazard estimation problems. Though the mean record age
has been analytically determined for random walk problems in earlier works \cite{maj1,wer},
there have been no results for the distribution of record age.

\begin{figure}
\includegraphics*[width=2.8in,angle=0]{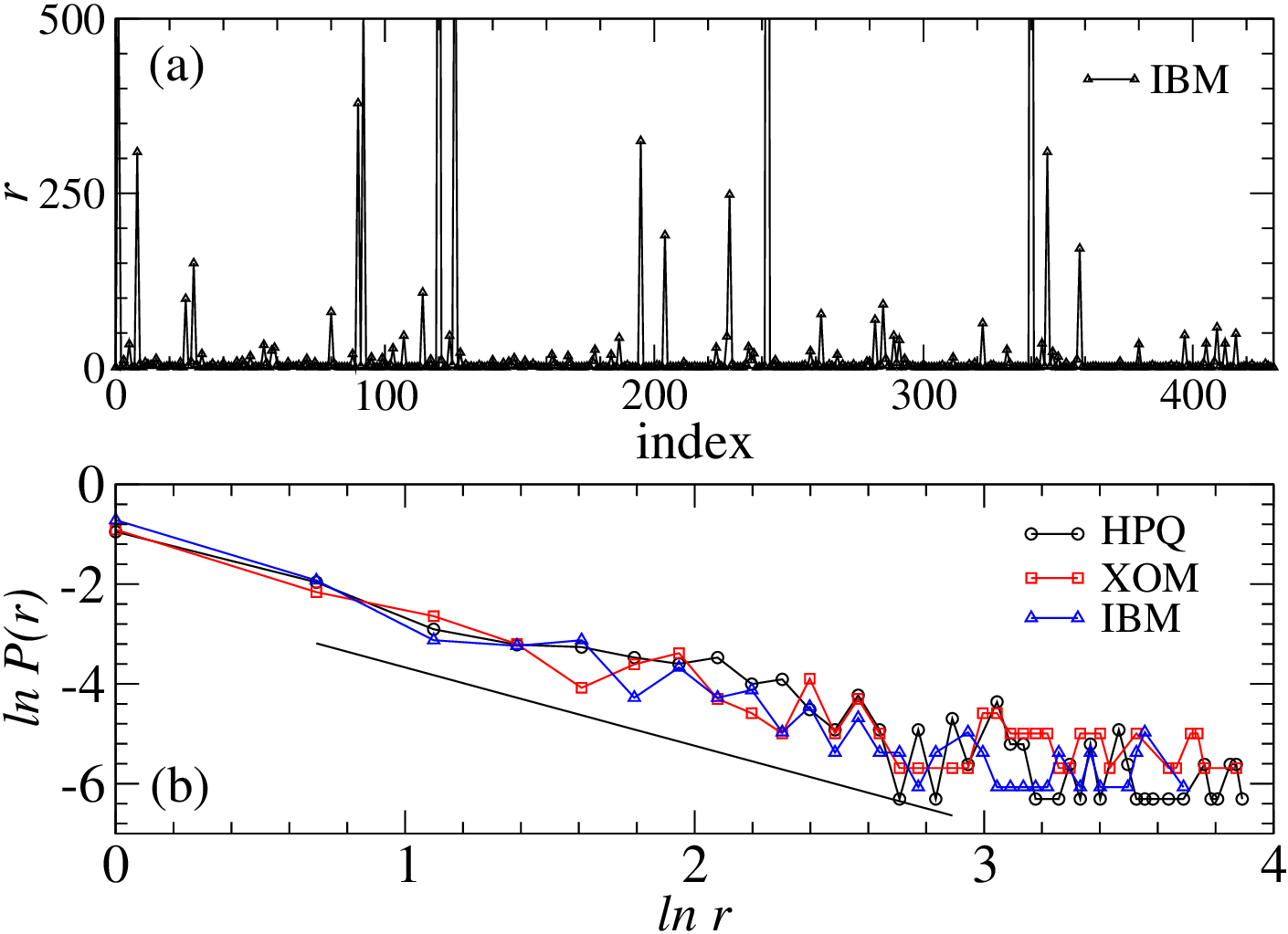}
\caption{(Color online) (a) Record ages (in days) calculated from IBM stock data.
(b) The distribution of record ages for three stocks. The best fit solid
line in (b) has slope $-1.58 \pm 0.15$.}
\label{fig2}
\end{figure}

\begin{figure}[t]
\includegraphics*[width=2.8in,angle=0]{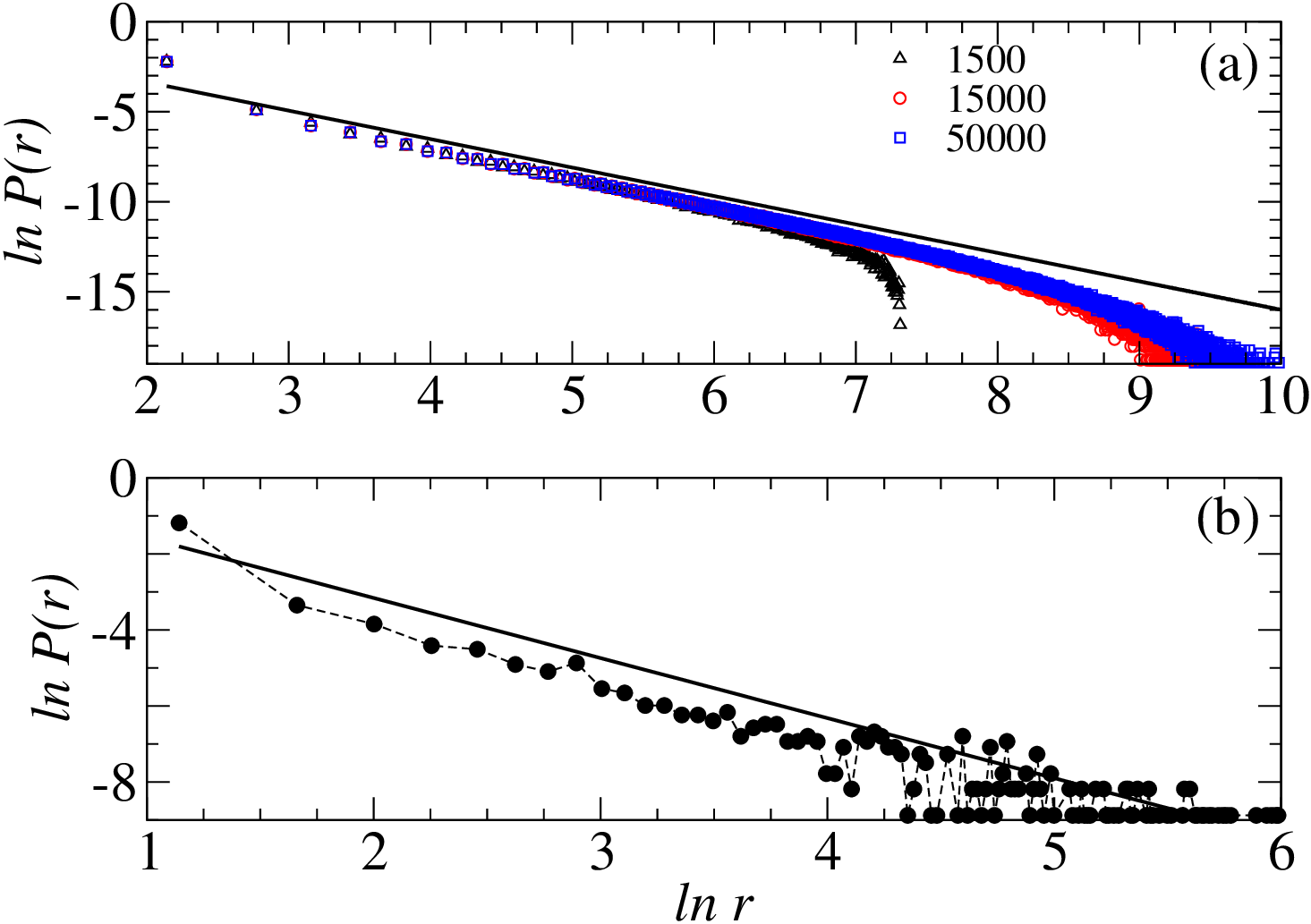}
\caption{(Color online) The distribution of record ages obtained from (a) GRW simulations
for three values of $N$ and (b) stock data other than those shown in Fig. \ref{fig2}(b).
See text for details. The solid line in (a) has slope $-1.652 \pm 0.006$ and in (b) has slope
$-1.611 \pm 0.051$.}
\label{alldata}
\end{figure}

In Fig. \ref{fig2}(a), we show the record ages obtained from
IBM stock data. In this, record ages longer than 500 are not
shown since they mask the details near $r=1$.  The longest record age
(not visible in Fig. \ref{fig2}(a) ) is 2313 days and the shortest is 
1 day. Thus, in this case, the record ages vary over 3 orders of magnitude.
Clearly, they depend on the length $N$ of data being considered since the longest
record age cannot exceed the length of data. Fig. \ref{fig2}(b) displays
the distribution of record age computed from the stock prices of three stocks
(HPQ, XOM and IBM)
with the longest available time series. In log-log plot shown in this
figure, the distribution, for most part, is consistent with a power law of the form
\begin{equation}
P(r) \sim A ~ r^{-\alpha}
\label{plaw}
\end{equation}
with the exponent $\alpha \approx 1.58$ and $A$ being the normalization
constant that can be written in terms of harmonic number $H_{N,\alpha}$. 
However, the tail of computed distribution flattens out due
to effect arising from the finite size of the data. In order to improve
the statistics, we use GRW simulations (Eq. \ref{grw}) with $\xi_i$ drawn from
normal distribution with parameters values
$\mu = \langle \mu_{emp} \rangle = 0.00031$ and $\sigma = \langle \sigma_{emp} \rangle = 0.015$.
These parameter values $\mu_{emp}$ and $\sigma_{emp}$ were computed from
the empirical stock data by averaging over the individual
values of $\mu$ and $\sigma$ obtained for each stock. The record age distribution
for each value of $N$, shown in Fig. \ref{alldata}(a), is averaged over $10^5$ GRW realizations.
Clearly, the distribution in Fig. \ref{alldata}(a)
can be represented as a power law in Eq. \ref{plaw} with the exponent $\alpha = 1.652 \pm 0.006$.
Significantly, it is independent of the value of $N$. In contrast to quantities like the
mean number of records which depend on $N$ \cite{maj1,wer,recs-rw1}, the distribution of record ages is
characteristic statistical property of record breaking events independent of the length
of data.
Further, as $N$ increases, the range over which the power law is valid also increases implying that
the tail behavior is a finite size effect. Within the parametric regime relevant for the stocks listed
in the Appendix, namely, $0.0001 \le \mu \le 0.0005$ and $0.01 \le \sigma \le 0.05$, we did not
find any systematic relation between the these parameters and the exponent $\alpha$.

Based on the results displayed in Fig. \ref{fig2}(b), we might expect that
all the individual stocks will display nearly the same value of $\alpha$ even
if $N$ is different for each one of them. Indeed, the value of the exponent lies
in the range $1.5 \le \alpha \le 1.8$ for the stocks listed in Appendix.
Hence, we combined the record ages computed from the rest of stock data
in Appendix (other than HPQ, XOM and IBM)
and the resulting distribution is displayed
in Fig. \ref{alldata}(b). The power law form (Eq. \ref{plaw}) is seen in the figure
with a value of exponent $\alpha \approx 1.611 \pm 0.051$.


\section{Longest record age}
Given that the record age is distributed as a power law, 
it is of interest to understand the distribution of longest record age. Clearly, shortest 
record age cannot be less than unity, a restriction arising from the resolution of
the data measurement. Similarly, any record age longer than the length of
the time series $N$ cannot be resolved. In ref. \cite{maj1,recs-rw1}, it
was pointed out that for a symmetric random walk process, the longest record age is proportional to $N$.
However, the distribution of longest record age has not been discussed earlier.

\begin{figure}[t]
\includegraphics*[width=2.2in,angle=0]{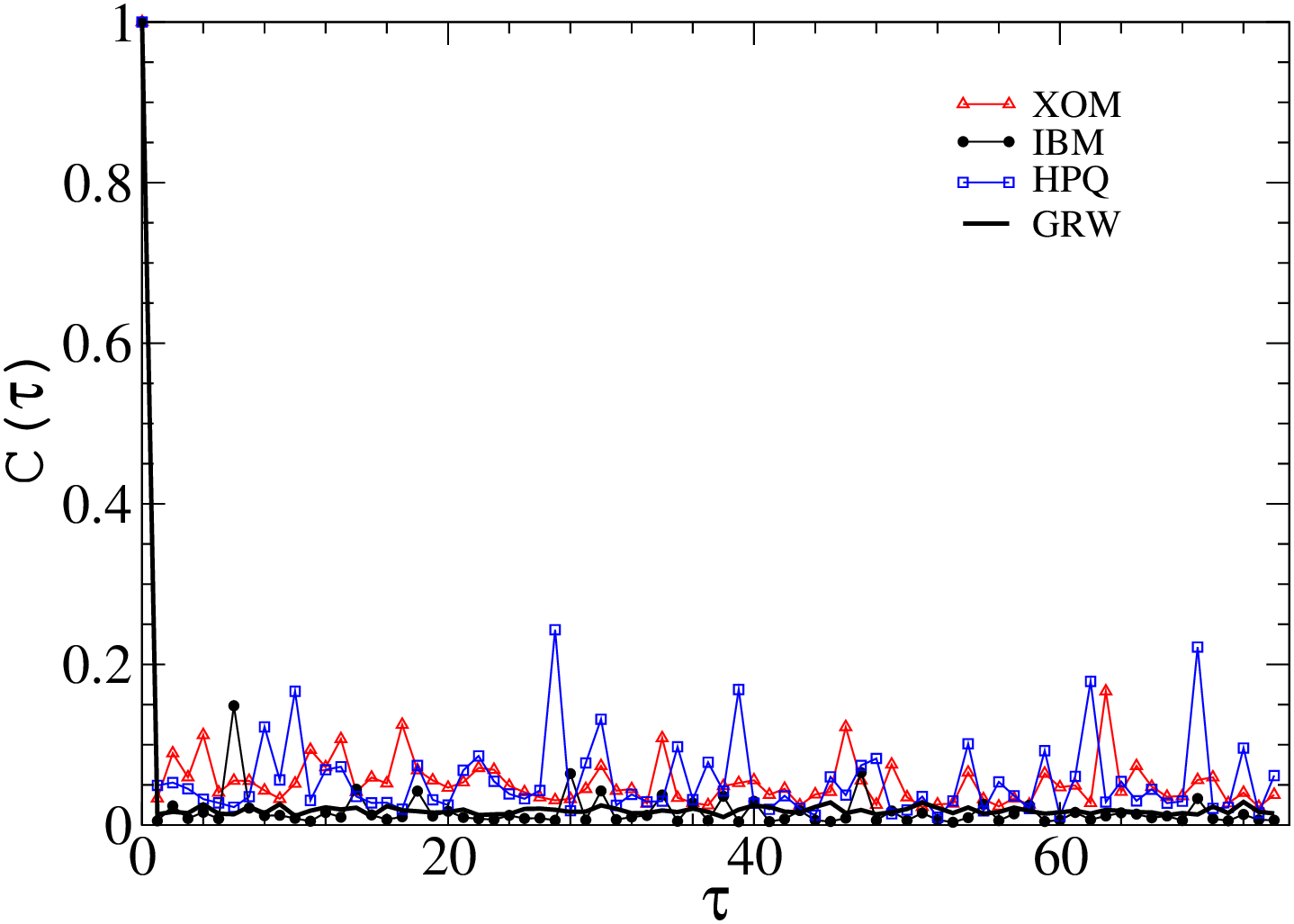}
\caption{(Color online) The autocorrelation function of the record ages obtained from three different
stock data and GRW simulations. For GRW results, we have used the same value
of $N$ as for IBM stock, $\langle \mu_{emp} \rangle = 0.00031$
and $\langle \sigma_{emp} \rangle = 0.015$. See text for details.}
\label{autocorr}
\end{figure}

In this section, we show that the longest record age falls in the class of
type-II generalized extreme value distribution, namely, the Fr\'{e}chet distribution \cite{gevt}. First clue
for this result arises from the record ages that are uncorrelated, to a good approximation.
Fig. \ref{autocorr} shows the autocorrelation function $C(\tau) = \langle x_t ~ x_{t+\tau} \rangle$ for
the stock data. It reveals that the record ages are, at best, weakly correlated. The record
ages obtained from GRW simulations (with parameters $\mu=\langle \mu_{emp} \rangle, 
\sigma=\langle \sigma_{emp} \rangle$ chosen similar to that in Fig. \ref{alldata}(a))
also show a similar behavior. For such fast decay of correlations, extreme value
theory for independent variables holds good \cite{lbr}. Hence, we can
expect the longest record age to follow the generalized extreme value distributions.

\begin{figure}[t]
\includegraphics*[width=2.8in,angle=0]{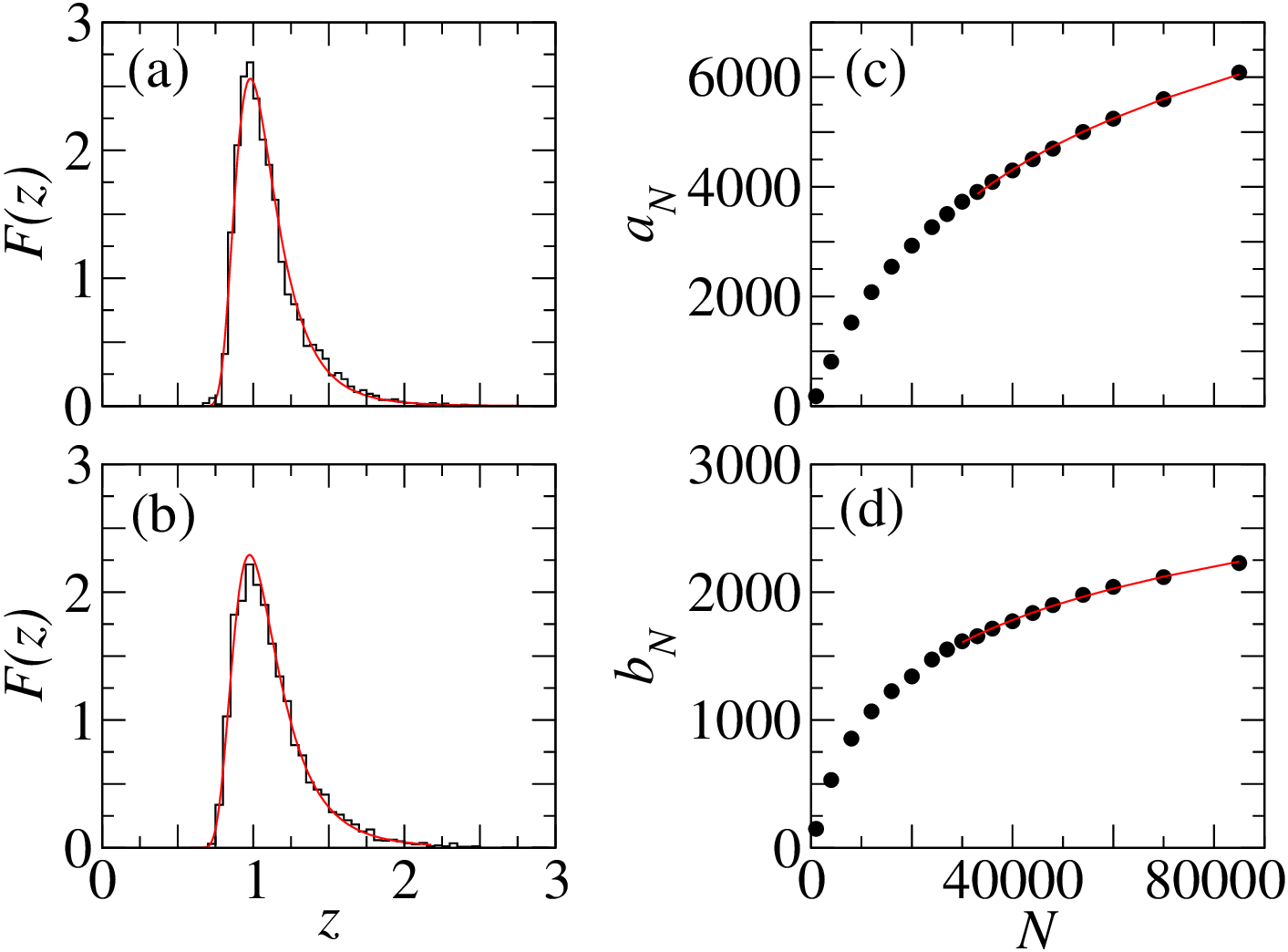}
\caption{(Color online) Scaled distribution of longest record age obtained from
GRW simulations with (a) $N=15000$ and (b) $N=85000$. The solid curve
is the Fr\'{e}chet distribution with shape parameter $k > 0$. (c) The location
parameter $a_N$ and (d) the scale parameter $b_N$ of the Fr\'{e}chet distribution
shown as a function of $N$. The solid line in (c,d) is the logarithmic fit
for $N>30000$.}
\label{xvt1}
\end{figure}

Fig. \ref{xvt1}(a,b) shows the distribution of longest record age $r_{max}$, in terms of
the scaled variable $z = 1+k(r_{max}-a_N)/b_N$, for the
GRW simulations with parameters same as for Fig. \ref{alldata}(a). In this, $a_N$ and $b_N$ are
location and scale parameters dependent on $N$. This figure reveals
a good agreement with the Fr\'{e}chet distribution \cite{gevt}
\begin{equation}
F(z) = \frac{1}{b_N} ~z^{-1-1/k} ~ e^{-z^{-1/k}}, \;\;\;\;\; (z>0),
\end{equation}
with shape parameter $k > 0$. 
This is the extreme value distribution consistent with results shown in
Figs. \ref{fig2},\ref{alldata}, {\it i.e}, the distribution of record ages $P(r)$ has a lower end
cut-off and its tail decays as a power law. The agreement with Fr\'{e}chet distribution
gets better for $N>>1$. The dependence of the location parameter $a_N$ and the scale 
parameter $b_N$ on $N$ shown in Fig. \ref{xvt1}(c,d) reveals that $\ln N$ function
provides a good representation of the data for $N > 30000$. Using this fit and the mean of Fr\'{e}chet distribution
$\langle z \rangle = a_N + (b_N/k) (\Gamma(1-k)-1)$,
we get the asymptotic  mean of the longest record ages as $\langle r_{max} \rangle \propto \ln N$.
This is the result obtained analytically in Ref. \cite{recs-rw1} without using extreme value theory.

\begin{figure}
\includegraphics*[width=2.2in,angle=0]{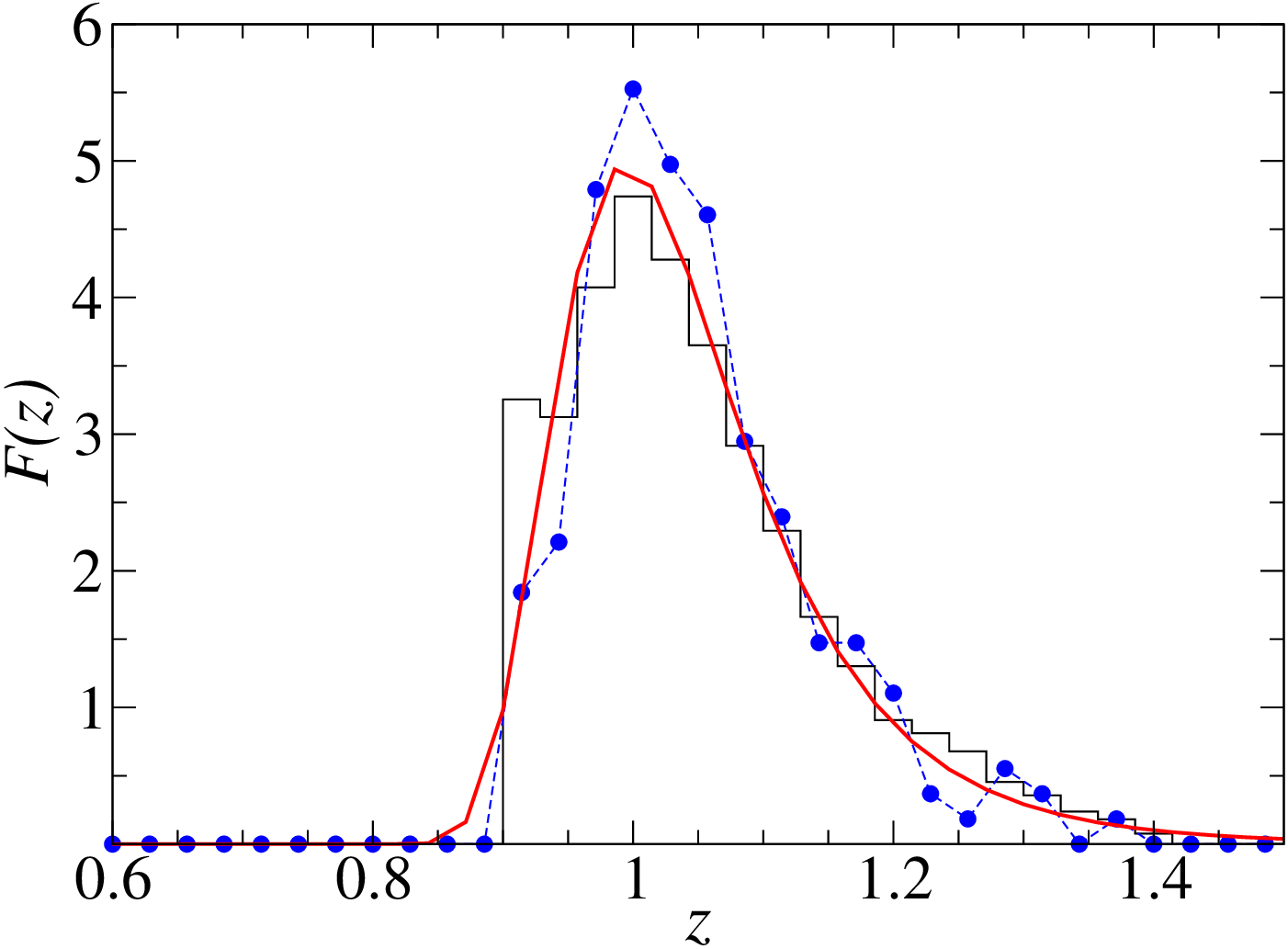}
\caption{(Color online) The distribution of scaled longest record ages computed
from stock data (solid circles), GRW simulations (histogram) with $N=1000$. The solid
curve is the Fr\'{e}chet distribution with parameters $a_N$ and $b_N$ corresponding to $N=1000$.}
\label{xvt2}
\end{figure}

Finally, we compute the distribution of longest record ages from the stock data.
To circumvent the shortage of data, we divided the empirical stock data into windows
of length $N=1000$. The longest record age from each of these windows was tabulated
for each stock. All such data of extreme record ages from all the stocks were combined
together to compute the (scaled) distribution shown in Fig. \ref{xvt2} as solid circles.
The histogram in the figure is obtained from $10^5$ ensemble GRW simulations with $N=1000$
and other parameters chosen as done in Fig. \ref{alldata}.
The solid curve is the Fr\'{e}chet distribution with $k > 0$.  The distribution $F(z)$ computed from
stock data displays a reasonable agreement with Fr\'{e}chet distribution. The deviations could
partly be attributed to the insufficient stock market data to compute extreme record ages.
We must also point out that both GRW simulations and stock data display 
pronounced deviation from Fr\'{e}chet distribution for $z<1$.

\section{summary and discussions}
In summary, we have analyzed the stock data for two quantities of interest in the study of
record statistics, namely, the distribution of record ages and the longest record age.
The results have been obtained based on the analysis of 18 stocks for which the
data is available in the public domain. We also study the geometric random walk series
as a suitable reference model in the context of the time series of stocks. 
For the stock data and the GRW simulations, the record ages are distributed as a power law
with exponent in the range $1.5 \le \alpha \le 1.8$. The record ages are uncorrelated, to
a good approximation. The longest record ages are well described by the Fr\'{e}chet distribution of
the extreme value theory.

The results presented in this work also applies to the records statistics of the positions
of a standard random walker. This is possible because the random walk
and GRW are related through a simple time-independent transformation.
The record age
distribution $P(r)$ is independent of $N$ to within the numerical errors and it does
not preclude the mean record age from being dependent on $N$ \cite{recs-rw1}.
Record ages being
nearly uncorrelated implies that predicting the length of time before the occurrence of
next record event based on historical data is unlikely to be easy even though the mean
record age can be determined \cite{maj1}.
The longest record age is Fr\'{e}chet distributed for $N>>1$ and pronounced deviations
exist for small $N$.
While an analysis of longer and bigger portfolio of stock data will yield better
estimates for power law exponent $\alpha$ and also for the longest record age distribution,
it would be interesting to analytically obtain these results.

\appendix*
\section{Data used in the analysis}
In this work, we use the daily closing values, corrected
for splits and dividends, of the following stocks. These are
publicly accessible from {\tt finance.yahoo.com}. Standard stock symbols
are used to indicate stock names.

\begin{table}[h]
\centering
\begin{tabular}{|l|c|c|c|}
\hline
Stock & Years & Length  & Stock \\
      &       & of data & Exchange        \\
\hline
IBM ({\footnotesize IBM}) & 1962-2012 & 12764 & NYSE \\
GIS ({\footnotesize General Mills Inc.})&            1983-2012  &             7358    &       NYSE   \\
AAPL({\footnotesize Apple Inc.})  &          1984-2012  &            7067     &       NASDAQ   \\
XOM ({\footnotesize Exxon Mobil Inc.})  &          1970-2012  &             10777   &       NYSE   \\
FP.PA ({\footnotesize Total SA}) &          2000-2012  &             3435    &       PARIS   \\
GD   ({\footnotesize General Dynamics Co.})  &          1977-2012  &             8600    &       NYSE  \\
GE   ({\footnotesize General Electric Co.}) &          1962-2012  &             12764   &       NYSE   \\
HPQ  ({\footnotesize Hewlett-Packard Co.}) &          1962-2012  &             12747   &       NYSE   \\
NTT  ({\footnotesize Nippon Telegraph ...}) &          1994-2013  &             4656    &       NYSE   \\
SNP  ({\footnotesize China Petroleum and ...}) &          2000-2013  &             3124    &       NYSE   \\
TM   ({\footnotesize Toyoto Motor Co.}) &          1993-2013  &             5021    &       NYSE   \\
VOW.DE ({\footnotesize Volkswagen AG}) &          2000-2013  &             3423    &       XETRA \\
CVX  ({\footnotesize Chevron Co.}) &          1970-2013  &             10910   &       NYSE   \\
WMT  ({\footnotesize Walmart Stores Inc.}) &          1972-2013  &             10238   &       NYSE   \\
F ({\footnotesize Ford Motors}) &          1977-2013  &             9141    &       NYSE   \\
COP  ({\footnotesize ConocoPhilips}) &          1982-2013  &             7878    &       NYSE   \\
BRK.A ({\footnotesize Berkshire Hathaway})&          1990-2013  &             5825    &       NYSE   \\
BP ({\footnotesize BP plc})   &          1988-2013  &             6445    &       NYSE   \\
\hline
\end{tabular}
\end{table}

\end{document}